\documentclass[journal]{IEEEtran}
\usepackage{url}
\usepackage{graphicx}
\usepackage{hyperref}
\usepackage{balance}
\usepackage{makecell}
\usepackage{stfloats}
\usepackage{color,soul}
\usepackage{amsmath}
\usepackage{multirow}
\usepackage{hhline}
\usepackage{cite}
\usepackage{amssymb}
\usepackage{amsthm}
\usepackage{mathtools}
\usepackage{multicol}
\usepackage{comment}
\usepackage{adjustbox}
\usepackage{multirow}
\usepackage{dirtytalk}

\begin{document}

%\sptitle{SI: Industry 4.0: Computing, Communications, and Sensing}

%\title{Digital Humans and Haptic Interfaces for Metaverse Applications: A Comprehensive Survey}
\title{MetaDigiHuman: Haptic Interfaces for Digital Humans in Metaverse} 
\author{Senthil Kumar Jagatheesaperumal, Praveen Sathikumar, and Harikrishnan Rajan

\thanks{S.K.Jagatheesaperumal, P.Sathikumar and H. Rajan, are with Department of Electronics \& Communication Engineering, Mepco Schlenk Engineering College, Sivakasi, (Tamilnadu), 626005 India. (e-mail: senthilkumarj@mepcoeng.ac.in, praveensathi2003\_ec@mepcoeng.ac.in
rajanappumon\_ec@mepcoeng.ac.in )}
}

\maketitle

\begin{abstract}
The way we engage with digital spaces and the digital world has undergone rapid changes in recent years, largely due to the emergence of the Metaverse. As technology continues to advance, the demand for sophisticated and immersive interfaces to interact with the Metaverse has become increasingly crucial. Haptic interfaces have been developed to meet this need and provide users with tactile feedback and realistic touch sensations. These interfaces play a vital role in creating a more authentic and immersive experience within the Metaverse. This article introduces the concept of MetaDigiHuman, a groundbreaking framework that combines blended digital humans and haptic interfaces. By harnessing cutting-edge technologies, MetaDigiHuman enables seamless and immersive interaction within the Metaverse. Through this framework, users can simulate the sensation of touching, feeling, and interacting with digital beings as if they were physically present in the environments, offering a more compelling and immersive experience within the Metaverse.
\end{abstract}

\maketitle

\begin{IEEEkeywords}
Digital Humans, Metaverse, Haptic Interfaces, 3D Modelling, Virtual Reality, Augmented Reality.
\end{IEEEkeywords}

\vspace*{-5pt}

\section{Introduction}
The transformation of humans into virtual characters and their subsequent control by humans is made possible by the advancement of 3D modeling and animation technologies. The process involves creating animated cartoon-like models, capturing facial movements through motion capture, and employing real-time rendering for processing digital footage and achieving realistic graphics. The enhancement of digital humans through the creation of digital twins has emerged as a powerful tool for virtual-world interactions and real-world healthcare decision-making. This approach combines genomic data, artificial intelligence (AI), and extended reality technologies~\cite{surveswaran2023glimpse}. The ubiquity of digital or virtual humans is evident, and it is projected to be a significant trend in 2025. These virtual humans, facilitated by emerging technologies such as AI and real-time graphics, are set to revolutionize human-machine interactions. 

Haptic interfaces refer to devices that facilitate manual interaction with virtual environments or teleoperated remote systems. These interfaces offer a means to experience and appreciate three-dimensional objects while maintaining conservation standards. In haptic interfaces, the user remains in constant contact with the device through direct contact interfaces. Alternatively, intermediate contact interfaces limit user-device contact to situations when it is necessary. The primary function of a haptic interface is to provide synthetic stimulation to both proprioception (the sense of body position and movement) and skin sensation.

When designing haptic interfaces for aerospace flight control, it is crucial to take into account the demanding requirements of reliability and force density~\cite{begin2019design}. To meet these requirements, an effective force-reflecting haptic interface device should provide specific criteria that are validated through independent psychophysical testing. However, the degree to which each criterion can be fulfilled depends on the availability of actuators, sensors, materials, and computer technology. It is important to consider all three criteria simultaneously. For seamless interaction with the haptic interface, the Phantom device has been developed. This device enables precise force control by measuring the user's fingertip position. The Phantom haptic interface strives to strike a balance between the three aforementioned criteria, aiming to create an affordable and effective force-reflecting haptic interface using existing technologies. In the comprehensive study conducted by Promwongsa et al.~\cite{promwongsa2020comprehensive}, the aim is to enable haptic communications and facilitate the delivery of skill sets over a vast network. This goes beyond the conventional transfer of video or audio data typically associated with fixed networks.  

\begin{figure}[ht!]
 %\vspace{-1em}
  \centering \includegraphics[width=0.5\textwidth]{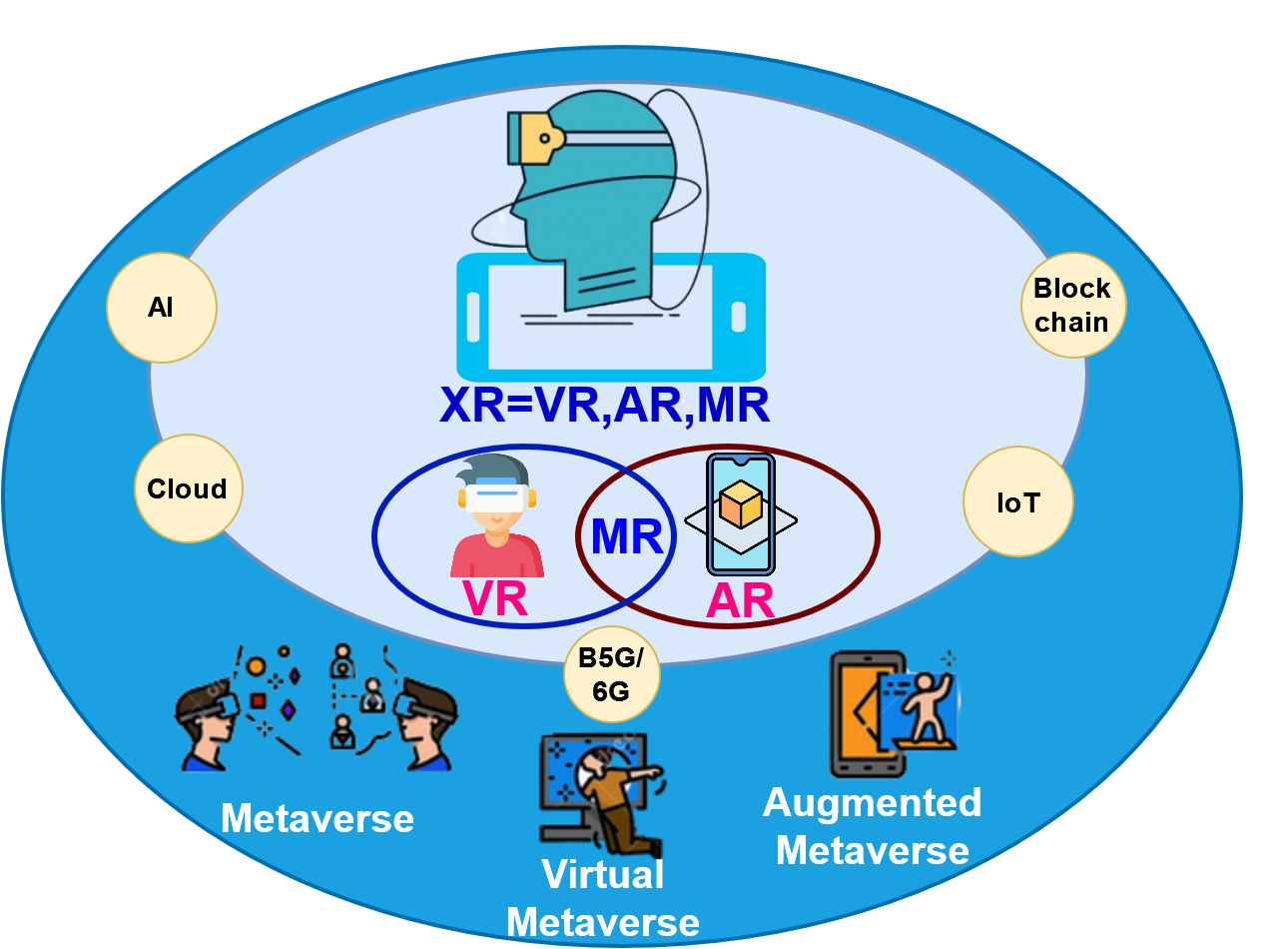}
  \caption{Major supporting components of Metaverse and its categories.}
   \label{fig:meta1}
   %\vspace{-1em}
\end{figure}

The concept of the metaverse envisions a hypothetical iteration of the Internet that transcends current limitations by creating a single, all-encompassing virtual world. This immersive environment is made possible through the utilization of virtual reality (VR) and augmented reality (AR) headsets. The application of the Metaverse extends to various domains in the real world, including tourism, education, manufacturing, banking services, healthcare, and more, as outlined in~\cite{koohang2023shaping}. In these sectors, the Metaverse holds significant potential for transformative experiences and innovative solutions. The core integrated components of the Metaverse in association with the enabling technologies are shown in Fig~\ref{fig:meta1} together to provide a rich and immersive experience for the users.
%------------------------------------------------------------------------------
%------------------------------------------------------------------------------

\textcolor{black}{Advanced sensors and actuators enable devices, such as IoT-enabled haptic gloves and suits, could be used to track human motion in real time. Comprehensive motion tracking is made possible by the haptic suits, which track larger body movements while the haptic gloves record fine motor movements. Across the user's body, a range of actuators could be configured to replicate different touch sensations to offer distributed tactile input. Advanced collision detection techniques through dynamic simulations via VR/AR gadgets could provide realistic user-digital environment interaction. To improve the experience with immersion and realism, precise rendering algorithms could be fine-tuned to provide accurate force feedback in Response to the user's interactions with virtual objects. 
}

In this article, we consider the setting of integrating the haptics with digital humans for Metaverse platforms with the consideration of AI and modeling as well as interfacing platforms. Specifically, we emphasize the role of various tools and platforms while fine-tuning the ethical constraints in digital humans on the remaining part.

\section{DIGITAL HUMANS IN INTELLIGENT SYSTEMS}

\subsection{Haptics: Present and Future}

Haptics play a significant role in enhancing the immersive experience of virtual reality environments. The application of haptics extends across various devices, ranging from wearables, handhelds, encountered-type devices, and props to mid-air approaches, as noted by Wee et al.\cite{wee2021haptic}.  Haptic interfaces have demonstrated successful utilization in diverse task domains. These include telesurgery, medical and surgical VR systems, rehabilitation, tele-and micromanipulation, telemaintenance, virtual prototyping, scientific visualization, and education~\cite{abdi2020haptics}. Despite these achievements, the full scope of their application potential remains largely unexplored and underutilized. Another notable advancement in this field is the AmbientTransfer system proposed by Li et al.~\cite{li2022ambienttransfer}. This system maps varying levels of haptic feedback to different body parts of users. The incorporation of AmbientTransfer can significantly enhance user immersion and presence while engaging with immersive videos.

In the present era, kinesthetic haptic interfaces have become commercially available for a wide array of applications. To ensure optimal performance, a range of highly specialized devices has been developed. For instance, the Laparoscopic Impulse Engine by Immersion exhibits a kinematic design and output capabilities that precisely meet the requirements of minimally invasive surgery simulation. Furthermore, with the aid of haptic feedback, users can manipulate micro-robots and establish seamless communication, as illustrated by Lee et al.\cite{lee2021real}, who introduced the use of haptic devices and a magnetic tweezer system.

These authors in~\cite{zhou2020calming} proposed a mechanism to reduce the anxiety of patients by experiencing their heartbeat-like vibration to be a calmer one through haptic interfaces. Robotic surgeries are used to transform the current surgery method more easily and flexibly. Haptic information displays differ from autonomous robots and teleoperators used in surgical applications and have different requirements than the haptic devices designed for the training and simulation of surgical procedures. The primary mode of sensory feedback has been through 3-D visual observation using stereo endoscopes. A fully actuated exoskeleton that covered the degrees of freedom of the human arm while providing enough range of motion, speed, strength, and haptic-rendering function to provide neuro-rehabilitation therapy to severely and mildly affected patients.~\cite{zimmermann2022performing}

The research conducted by Shi et al.~\cite{shi2022multiscale} focuses on utilizing graph convolutional networks (GCNs) to recognize human emotions. Specifically, they employ a multiscale approach that separates spatial modeling into distinct processes. In the work outlined in~\cite{macdonald2021user}, the objective is to create calming affective haptic stimuli by eliciting user preferences within various social networks. Here, the haptic experiences promote a sense of relaxation and emotional well-being, considering the diverse preferences and characteristics of users within different social contexts.

\subsection{The Role and Impact over Metaverse}
Much like AI, the metaverse combines various elements of social media to form a persistent three-dimensional realm where users are represented by avatars, as digital humans, which has been already popular in online video games, exemplifying its potential. They enable the testing and refinement of AI algorithms, helping to create more advanced and sophisticated technologies. As noted by Li et al.~\cite{li2023simulation}, simulations with digital humans have become essential in the pursuit of state-of-the-art AI systems. The convergence of AR, AR, and social networking has set the stage for the emergence of the metaverse. With the continuous advancement of technology and the growing interest in immersive experiences, the realization of a fully functional metaverse is an exciting prospect for the future. \textcolor{black}{The creation of VR/AR and haptic interfaces by considering people with disabilities into account encompasses features including customized interfaces, alternate input methods, and adjustable feedback intensity. MetaDigiHuman hopes to give all users, regardless of physical ability, an equal and engaging experience by implementing these accessibility features.}

The concept of the DeMetaverse~\cite{wang2022metaverses} introduces a novel approach to metaverse development. It revolves around the idea of a decentralized autonomous metaverse, enabled by Decentralized Autonomous Organization (DAO) technology. In this context, the content and games created within the DeMetaverse remain the property of the users who create them. 

\textcolor{black}{Distributed computing, load balancing, and improved data processing are some of the techniques used to guarantee seamless performance with a large number of concurrent users. For the system to continue operating effectively and responsively in the Metaverse, these real-time scalable measures are essential.
}

\section{MODELLING AND INTERFACE PLATFORMS}
\label{sec:model}
\begin{figure*}[htp]
    \centering
    \includegraphics[width=15cm]{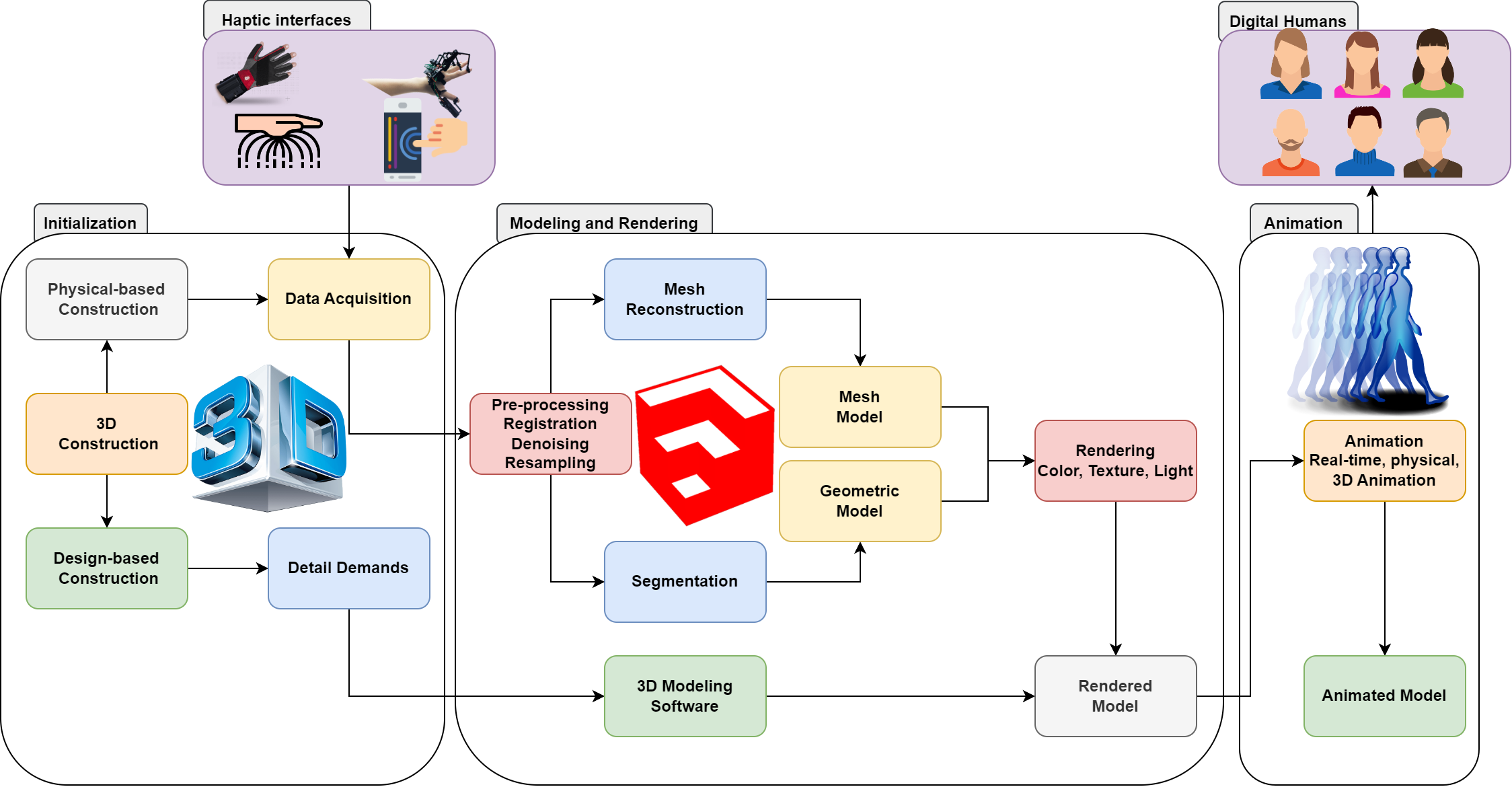}
    \caption{Haptic interfaces used by the modeling and rendering tools to interface with the digital humans in the Metaverse platforms}
    \label{fig:1}
\end{figure*}

Digital Human Models (DHM), also referred to as three-dimensional manikins, serve as virtual representations of humans and play a vital role in designing physical prototypes with precision and efficiency. DHM software programs, such as Jack, Ramsis, and Safework, leverage technology similar to Computer-Aided Design (CAD) programs. These programs enable users to import their 3D CAD models into a virtual environment, where DHMs of different sizes can be integrated alongside the model for comprehensive design analysis. Figure~\ref{fig:1} shows the sequence of stages involved from the haptic interfaces to the digital humans driven with the aid of robust modeling and rendering tools.

\textcolor{black}{IoT-enabled haptic gloves integrated with sophisticated sensors and actuators for high-fidelity tactile feedback and real-time motion tracking are among the main hardware units. A digital human animation module, motion capture algorithms, and a haptic rendering engine make up the software architecture. Together, these elements guarantee accurate tracking of movements and authentic tactile experiences. Using sophisticated tracking algorithms to reduce latency and guarantee precise mapping of human actions onto digital avatars, the integration process synchronizes hardware and software. By improving real-time feedback methods and optimizing algorithms, issues like motion tracking accuracy and latency reduction could be resolved. 
}

In the realm of human-robot interaction, the research presented by Lestingi et al.~\cite{lestingi2022model} introduces an innovative framework that harnesses the power of modeling, verification, and learning techniques. This framework aims to facilitate seamless human interaction with robots while ensuring the preservation of human well-being and the successful completion of missions, even in the face of uncertainties. By employing robust mathematical guarantees, this approach provides a solid foundation for maintaining human safety and achieving mission objectives throughout the interaction. By combining the capabilities of DHMs with advanced modeling, verification, and learning techniques, this research contributes to the advancement of human-robot collaboration. 

\begin{itemize}
    \item \textit{Haptic Feedback Devices:} IoT-enabled haptic gloves or haptic suits, play a pivotal role in elevating the user experience within the Metaverse. These devices have the remarkable ability to deliver tactile feedback and realistic touch sensations, enabling users to feel a genuine sense of physical interaction with digital humans. By simulating the sense of touch, haptic feedback devices further enrich the immersive environment of the Metaverse.

    \item \textit{VR and AR Platforms:} With Oculus Rift, HTC Vive, and Microsoft HoloLens, being some of the popular VR/AR platforms, they serve as immersive gateways for engaging with digital humans within the expansive Metaverse. By leveraging these advanced platforms, users are transported into a first-person perspective, enabling them to perceive, interact, and engage with digital humans and their surroundings profoundly and realistically. 

    \item \textit{3D Modeling Software:} Software tools such as Blender, Maya, and 3ds Max are widely utilized for the creation of intricate 3D models of digital humans. These advanced platforms empower artists and designers to meticulously sculpt, texture, and animate lifelike human characters, allowing for seamless integration into the immersive Metaverse environment. 

    \item \textit{Motion Capture Systems:} Motion capture technology, exemplified by %industry-leading solutions like OptiTrack and Vicon, proves invaluable in capturing the intricate movements and gestures of real individuals and seamlessly transferring them to digital characters. By employing this technology, lifelike animations can be achieved, significantly enhancing the realism and authenticity of digital humans within the immersive realm of the Metaverse. 
    \textcolor{black}{platforms such as Oculus Rift, HTC Vive, and Microsoft HoloLens could be used to improve the integration of current VR and AR technologies with the MetaDigiHuman architecture. Users can enter immersive environments with these VR/AR devices, which provide realistic interactions with virtual individuals and first-person perspectives. To produce a smooth and engaging experience, the integration process makes use of their sophisticated tracking capabilities and high-resolution screens. }
    
    \item \textit{Game Engines:} Game engines such as Unity or Unreal Engine stand as robust and versatile tools for crafting interactive virtual environments and seamlessly integrating digital humans with haptic interfaces. These engines offer a wide range of capabilities, including real-time rendering, physics simulations, and advanced scripting systems, empowering creators to breathe life into the Metaverse. 

\end{itemize}
With these tools, creators can bring their artistic vision to life, ensuring that digital humans exhibit a high level of realism and visual fidelity within the Metaverse, creating a heightened level of realism and engagement.

\textcolor{black}{IoT-enabled haptic gloves suits with a network of high-precision sensors included in the gloves and outfits to track the mobility of human joints in real-time within the MetaDigiHuman framework. To precisely map the movements onto digital human avatars, sophisticated algorithms process the detailed motion data that the sensors record. The haptic suits are fitted with a range of actuators that simulate touch over the user's body by applying different pressure and vibration patterns to the skin to provide dispersed tactile input. Advanced collision detection and force rendering methods could enable dynamic simulations between virtual persons and controlled objects. To ascertain how the user's avatar interacts with virtual objects, the collision detection system may employ spatial analysis, making sure that contact points are accurately identified. 
}

\textcolor{black}{MetaDigiHuman systems in controlled environments could be configured to perform various tasks that require interaction with digital humans and haptic interfaces. Empirical evidence from the presented studies showcases performance metrics such as Response time, accuracy of motion tracking, and the fidelity of force feedback. Based on the comparative analyses with the traditional interaction methods, demonstrating significant improvements in user experience and interaction quality could be easily tailored for the MetaDigiHuman framework. 
}

\vspace*{-5pt} 
\section{ETHICAL CONSIDERATIONS ON DIGITAL HUMAN}
~\label{sec:ethical}
Ethical considerations surrounding the use of haptic interfaces in metaverse environments typically encompass concerns regarding privacy, consent, and the potential for haptic technology to be misused. A digital human can provide users with data such as biometrics and human health records. Governments can establish regulations and standards to ensure transparency and responsible use of digital humans, addressing issues like privacy breaches. Professional organizations can develop ethical guidelines outlining the best practices for employing digital humans in their respective fields. These guidelines serve as a framework to guarantee the ethical and responsible utilization of digital humans. Developers can actively work towards identifying and eliminating biases in digital humans by utilizing diverse datasets, testing for bias and discrimination, and involving diverse teams in the development process to create more realistic digital humans. 

%------------------------------------------------------------------------------
%\begin{figure}[ht!]
 %\vspace{-1em}
  %\centering \includegraphics[width=0.5\textwidth]{figures/Challenges.png}
  %\caption{Some of the popular challenges in the proposed MetaDigiHuman framework.}
   %\label{fig:chal}
   %\vspace{-1em}
%\end{figure}
%------------------------------------------------------------------------------

With stringent ethical considerations in place, the integration of digital humans and haptic interfaces holds potential applications in the fields of the Metaverse, spatial computing, and Brain-Computer Interfaces (BCIs). Spatial computing enables the control of computing devices through natural gestures and speech, while BCIs allow for communication with computing devices solely through brain activity. BCIs can be used for controlling synthetic limbs or empowering individuals with paralysis to operate computers. In the context of Cyber-Physical-Social Systems (CPSS), meta-sensing, as proposed in~\cite{liu2022metasensing}, offers a solution for smart sensing of human factors. This is particularly important due to the inadequate consideration of human factors, such as digital twins, in CPSS environments. \textcolor{black}{In protecting consent and privacy in digital human interactions, it is suggested to implement safeguards including user permission procedures, safe data encryption, and moral principles to avoid abuse and preserve user privacy. Encouraging a secure and ethically robust immersive environment in the Metaverse demands safety measures.}

In scenarios where the human operator is not present within the system, it becomes vital for the system to identify errors and devise strategies to overcome them. In this context, it is imperative for the human operator, who oversees the system, to make well-informed decisions supported by logical explanations.%~\cite{li2020explanations}.
By employing digital humans driven by the Metaverse in an ethically conscious setting, such supervision can effectively address potential issues and ensure efficient management. \textcolor{black}{The compatibility challenges exist in the framework, where it can be used with well-known environments like Unreal Engine and Unity, where issues like data synchronization and API compatibility are essential. 
}

\textcolor{black}{Prospective future research directions and particular applications for the MetaDigiHuman frameworks, could be in highlighting the framework's revolutionary potential in a range of industries. With its accurate haptic feedback and real-time simulations, MetaDigiHuman can be used in the healthcare industry for remote surgical training and rehabilitation, enhancing both the abilities of practitioners and the outcomes of their patients. Through tactile interactions, the framework can be used in education to build immersive learning environments where students interact with virtual individuals in virtual classrooms, improving engagement and retention. With MetaDigiHuman, the entertainment sector can create more realistic and dynamic gaming environments where users can interact in real-time with virtual people and places. Subsequent investigations will concentrate on enhancing the haptic feedback mechanisms, broadening the scope of sensory inputs, and guaranteeing a smooth merge with developing VR/AR technologies.
}

\section{CONCLUSIONS AND FUTURE STUDIES} 
\label{sec:conclusion}
In this article, we present an innovative platform called MetaDigiHuman, which combines the elements of the Metaverse, digital humans, and haptic interfaces to create a unique digital space experience. We aim to explore the possibilities of MetaDigiHuman and unlock a new dimension of digital engagement within the Metaverse by leveraging haptic interfaces. We delve into the strategies required to harness the potential of MetaDigiHuman and emphasize the significance of up-to-date modeling and interfacing platforms that enable effective communication and interaction among digital humans. Additionally, we address the ethical considerations involved in facilitating seamless and immersive experiences for digital humans within the Metaverse. The transformative nature of MetaDigiHuman holds immense promise in revolutionizing our digital interactions and reshaping our engagement in the Metaverse. In the future, we are eager to further explore the capabilities and implications of this groundbreaking framework and its potential impact on the digital realm.

%\section{ACKNOWLEDGMENTS}
%We thank the Management, Principal, and the Head of the Department of Mepco Schlenk Engineering College, Sivakasi, India for extending necessary motivation and advice in this research. This work is supported in part by the Centre for Robotics and Embedded Systems, and by the Mepco Incubation Centre-Sivakasi, India.

%A whole-body sensation in humans through wireless connections is mostly used in medicine and gaming~\cite{jung2022wireless}. This will convey navigation instructions, translate musical tracks into tactile patterns, and support sensory replacement feedback for the control of robots.

\def\refname{REFERENCES}

\begin{IEEEbiographynophoto} {Senthil Kumar Jagatheesaperumal} is currently an associate professor (senior grade) with the Department of Electronics and Communication Engineering, Mepco Schlenk Engineering College, Sivakasi, 626005, India. His research areas include IoT, robotics, embedded systems, and wireless communication. He received his Ph.D. in Information and Communication Engineering from Anna University, Chennai, India. He is a Life Member of IETE and ISTE. Contact him at senthilkumarj@mepcoeng.ac.in.
\end{IEEEbiographynophoto}

\begin{IEEEbiographynophoto}{Praveen Sathikumar}{\,} is an undergraduate researcher at the Department of Electronics and Communication Engineering, Mepco Schlenk Engineering College, Sivakasi, 626005, India. His current research interests include IoT, Wireless Sensor Networks, and Artificial Intelligence. He received various awards and achievements in various project contests. He is a student member of The Institution of Electronics and Telecommunication Engineers (IETE). Contact him at praveensathi2003\_ec@mepcoeng.ac.in  %\vspace*{8pt}
\end{IEEEbiographynophoto}

\begin{IEEEbiographynophoto}{Harikrishnan Rajan} {\,} is an undergraduate researcher at the Department of Electronics and Communication Engineering, Mepco Schlenk Engineering College, Sivakasi, 626005, India. His current research interests include IoT, Wireless Sensor Networks, and Artificial Intelligence. He received various awards and achievements in various project contests. He is a student member of The Institution of Electronics and Telecommunication Engineers (IETE). Contact him at rajanappumon\_ec@mepcoeng.ac.in.
\end{IEEEbiographynophoto}

\end{document}